\documentstyle[12pt]{article}
\newcommand{\subs}[1]{\subsection{#1}\setcounter{equation}{0}}

\begin{document}\bigskip
\input epsf         
\hskip 3.7in\vbox{\baselineskip12pt
\hbox{NSF-ITP-96-01}\hbox{hep-th/9601038}}
\bigskip\bigskip\bigskip

\centerline{\large \bf Consistency Conditions for
Orientifolds and D-Manifolds}

\bigskip\bigskip

\centerline{\bf Eric G. Gimon}
\medskip
\centerline{Department of Physics}
\centerline{University of California Santa Barbara}
\centerline{Santa Barbara, CA\ \ 93106}
\centerline{e-mail: egimon@physics.ucsb.edu}
\bigskip
\centerline{\bf Joseph Polchinski} 
\medskip
\centerline{Institute for Theoretical Physics}
\centerline{University of California}
\centerline{Santa Barbara, CA\ \ 93106-4030}
\centerline{e-mail: joep@itp.ucsb.edu}
\bigskip\bigskip

\begin{abstract}
\baselineskip=16pt
We study superstrings with orientifold
projections and with generalized open string boundary conditions (D-branes).
We find two types of consistency condition, one related to the algebra of
Chan-Paton factors and the other to cancellation of divergences.  One
consequence is that the Dirichlet 5-branes of the Type I theory carry a
symplectic gauge group, as required by string duality.  As another application
we study the Type I theory on a $K3$ $Z_2$ orbifold, finding a family of
consistent theories with various unitary and symplectic subgroups of $U(16)
\times U(16)$.  We argue that the $K3$ orbifold with spin connection embedded
in gauge connection corresponds to an interacting conformal field theory
in the Type I theory.

\end{abstract}
\newpage
\baselineskip=18pt

\subs{Introduction}

One of the notable features of string duality has been the
convergence of many previously disjoint lines of development.  For
example, certain once-obscure string backgrounds, namely
orientifolds~[1-3] and D-manifolds~[3], have proven
to be dual to more familiar
backgrounds~[4-8].  In order to find
the nonperturbative structure underlying string duality it is
important to understand as fully as possible all limits of the
theory.  The purpose of the present paper is to develop the
consistency conditions for orientifolds and D-manifolds.

Orientifolds are generalized orbifolds.  In the orbifold
construction, discrete internal symmetries of the
world-sheet theory are gauged.  In
the orientifold, products of internal symmetries with world-sheet
parity reversal are also gauged.
D-manifolds are manifolds with special submanifolds (D-branes) on which
strings are allowed to end. These are labeled
by a generalized Chan-Paton index, each value of which corresponds to
restricting the string endpoint to a given submanifold of spacetime.

We will discuss consistency conditions of two types.  The first comes
from closure of the operator product expansion, which restricts the
action of the discrete gauge symmetries on the Chan-Paton index.  One
consequence is that D 5-branes in Type I string theory must have a symplectic
rather than orthogonal gauge projection: this is a world-sheet
derivation of a result previously found from string
duality~\cite{witinst}.  Also, D 3-branes and 7-branes are
inconsistent in the Type I superstring, while D 1-branes have an orthogonal
gauge projection.

The second condition is cancellation of divergences and anomalies at one
loop~\cite{gsdiv}, which can be recast in terms of consistency of the
field equations~\cite{cai}.  Here we focus on a simple example, the
Type I theory on a K3 orbifold.  We find all solutions to the
consistency conditions, leading to gauge groups which are various
unitary and symplectic subgroups of $U(16)
\times U(16)$.  Rather surprisingly, we do not find a solution with
the spin connection embedded in the gauge connection.  We argue that
this theory, while it must exist, does not correspond to a free
conformal field theory.  Finally, we discuss various related work. 

\subs{Orientifolds and Chan-Paton Factors}

The orientifold group contains elements of two kinds.  The first are
purely internal symmetries $g$ of the world-sheet theory, forming a
subgroup $G_1$.  For the purposes of the present paper we will think
of these as spacetime symmetries, though more generally (as in
asymmetric orbifolds) one could consider symmetries whose
spacetime interpretation is less clear.  The second are elements of
the form $\Omega h$, where $\Omega$ is the world-sheet parity
transformation and $h$ is again a spacetime symmetry, now chosen from
a set $G_2$.  Closure implies that $\Omega h \Omega h' \in G_1$ for
$h, h' \in G_2$, and if all elements of $G_2$ commute with $\Omega$
this is simply $G_2 G_2 = G_1$.  The full orientifold group is $G = G_1
+ \Omega G_2$.

In the orientifold construction this group is gauged, meaning that one
sums over all group elements around any nontrivial path on the
world-sheet.  This projects onto states invariant under $G_1$ and
$\Omega G_2$.  Elements of $G_1$ also lead to twisted closed strings,
from a gauge transformation in going around a closed string.  The
factor
$\Omega G_2$ means that orientation-reversal (combined with a $G_2$
action on the fields) is now part of the local symmetry group, so that
unoriented world-sheets are included.  The elements of $\Omega G_2$
do not give rise directly to new (twisted) sectors of the string
Hilbert space; we will discuss later the extent to which it is useful
to think of the open strings as being these twisted states.

The Chan-Paton index $i$
labels a set of a submanifolds (D-branes)
$M_i$, with a string end-point in state $i$ constrained to lie on
$M_i$.   Some of the $M_i$ may be coincident.
Each element of the discrete gauge group will have some action on the
Chan-Paton index.   Denote a general open string state by
$|\psi,ij\rangle$, where
$\psi$ is the state of the world-sheet fields and $i$ and $j$ are
the Chan-Paton states of the left and right endpoints; the boundary
conditions on the fields in $\psi$ are of course $i,j$-dependent.  The elements
$g$ act on this as
\begin{equation}
g:\qquad |\psi,ij\rangle \ \to\
(\gamma_g)_{ii'}|g\cdot\psi,i'j'\rangle
(\gamma_g^{-1})_{j'j}.  \label{gact}
\end{equation}
for some matrix $\gamma_g$ associated with $g$.\footnote
{Some time after the completion of this paper, we learned that much of
the following formalism 
was developed for $Z_2$ orientifolds of the bosonic string by Pradisi
and Sagnotti in the early paper~\cite{PS}.}
 This form is
determined by the requirement that a general trace of products of
wavefunctions be invariant.  The action on the Chan-Paton factors must
also be consistent with the action on the fields.  That is, for each
D-brane
$M_i$, the spacetime-transformed D-brane $g\cdot M_i$ must
appear, and the only nonzero elements of
$\gamma_g$ are those connecting $M_i$ and $g\cdot M_i$.  If
$M_i$ is left fixed by $g$ then diagonal elements are
allowed.  Similarly, 
\begin{equation}
\Omega h:\qquad |\psi,ij\rangle \ \to\
(\gamma_{\Omega h})_{ii'}|\Omega h\cdot\psi,j'i'\rangle
(\gamma_{\Omega h}^{-1})_{j'j}.
\end{equation}
Note that the orientation reversal transposes the two endpoints.
The $\gamma_g$ and $\gamma_{\Omega h}$ are unitary.

To derive further constraints on the matrices $\gamma_g$ and $\gamma_{\Omega
h}$, let us first demonstrate that the discrete gauge group may not
include pure gauge twists, those with $g = 1 \in G_1$ with $\gamma_1$ nontrivial.
The point is that the allowed Chan-Paton wavefunctions must form a
complete set: the set of string wavefunctions $|\psi,ij\rangle$ must
include nontrivial states for all pairs $ij$.  One can see this
heuristically by noting that if there are states $ik$ and $jl$ for
some $k$ and $l$ (and therefore also $lj$ by CPT), then 
by a splitting-joining interaction one obtains also $ij$ and $lk$.
This interaction occurs in the interior of the string and so by
locality cannot depend on the values of the endpoints.  One can make
this precise by requiring that the annulus factorize correctly on the
closed string poles, so this is actually a one-loop condition---at
tree-level it would be consistent to truncate to block-diagonal
wavefunctions.  Now, if the identity appears in $G_1$, we have
the projection
\begin{equation}
|\psi,ij\rangle \ =
(\gamma_1)_{ii'}|\psi,i'j'\rangle
(\gamma_1^{-1})_{j'j}. 
\end{equation}
Since this holds for a complete set, Schur's lemma implies that
$\gamma_1 \propto 1$; we may as well set $\gamma_1 = 1$ because the
overall phase is irrelevant.

This implies a further restriction on the $\gamma_g$ and $\gamma_{\Omega h}$:
they must satisfy the algebra of the corresponding symmetries, up to a phase.
For example, $\gamma_{g_1} \gamma_{g_2} \gamma_{g_2^{-1} g_1^{-1}} \propto
1$, else we would contradict the result in the previous section.
As another example, suppose that $G_1$ includes an element of order
2, $g^2 = 1$.  Then on a string state,
\begin{equation}
g^2:\qquad |\psi,ij\rangle \ \to\
(\gamma_g^2)_{ii'}|\psi,i'j'\rangle
(\gamma_g^{-2})_{j'j},  \label{g2act}
\end{equation}
and so (by choice of phase)
\begin{equation}
\gamma_g^2 = 1.
\end{equation}
Similarly, if $G_2$ includes
an element of order 2, then $(\Omega h)^2$ acts as
\begin{equation}
(\Omega h)^2:\qquad |\psi,ij\rangle \ \to\
(\gamma_{\Omega h} (\gamma_{\Omega h}^T)^{-1})_{ii'}|\psi,i'j'\rangle
(\gamma_{\Omega h}^T \gamma_{\Omega h}^{-1})_{j'j}, \label{oh2}
\end{equation}
implying that
\begin{equation}
\gamma_{\Omega h}^T = \pm \gamma_{\Omega h}. \label{chitran}
\end{equation}

Let us apply this to the Type I theory.
The Type I theory is an orientifold of the Type IIB theory with the
single nontrivial element $\Omega$; that is, $h = h^2 = 1$.
Tadpole cancellation, to be reviewed in the next section, requires
that the orientifolding be accompanied by the inclusion of $n$ 9-branes,
corresponding to purely Neumann boundary conditions.
If $\gamma_{\Omega}$ is symmetric, we can choose a basis such that
$\gamma_{\Omega} = I$.  If $\gamma_{\Omega}$ is antisymmetric, we can choose
a basis such that
$\gamma_{\Omega}$ is the symplectic matrix
\begin{equation}
M = \left[ \begin{array}{cc} 0&iI\\-iI&0 \end{array} \right],
\end{equation}
where $n$ must be even.
For the massless open string vector, the $\Omega$ eigenvalue of the
oscillator state is $-1$.  For $\gamma_{\Omega}$ symmetric, the Chan-Paton
wavefunction of the vector is then antisymmetric, giving the gauge
group $SO(n)$.  For $\gamma_{\Omega}$ antisymmetric, the massless vectors
form the adjoint of $USp(n)$.  Tadpole cancellation requires $SO(32)$.

Now let us consider adding 5-branes.  The Chan-Paton index runs
over both 9-branes and 5-branes.  The only freedom in
eq.~(\ref{chitran}) is the overall sign.  Since we are required to
take the $SO$ projection on the 9-branes it appears that we are
required to take the same projection on the 5-branes.  This is in
contradiction with ref.~\cite{witinst}, where it was found that
string duality requires a symplectic gauge group on the Type~I
5-brane.  To understand this we need to be somewhat more careful.

The point is that, although $\Omega^2$ acts trivially on the
world-sheet fields, it may be a nontrivial phase in various sectors
of the Hilbert space.  The phase of $\Omega$ is determined by the
requirement that it be conserved by the operator product of the
corresponding vertex operators.  Thus, the massless vector
state, with vertex operator $\partial_t X^\mu$, necessarily has
$\Omega = -1$ because $\Omega$ changes the orientation of the tangent
derivative $\partial_t$; we have used this fact two paragraphs
previously.  In the 55 sector (that is, strings with both ends on a
5-brane), for the massless vertex operator is $\partial_t X^\mu$
($\Omega = -1$) for $\mu$ parallel to the 5-brane, and $\partial_n X^\mu$
($\Omega = +1$) for $\mu$ perpendicular. 
On these states, $\Omega^2 = 1$, and the same is true for
the rest of the 99 and 55 Hilbert spaces.  To see this, use the fact
that $\Omega$ multiplies any mode operator
$\psi_r$ by $\pm e^{i\pi r}$.  (Details of the mode expansions are
given in section 3.3.)  In the Neveu-Schwarz sector this is $\pm i$,
but the GSO projection requires that these modes operators act in
pairs.\footnote{The OPE is single-valued only for GSO-projected vertex
operators.}  So $\Omega = \pm 1$, and this holds in the R sector as
well by supersymmetry.

Now consider the Neveu-Schwarz 59 sector.  The four $X^\mu$ with mixed
Neumann-Dirichlet boundary conditions, say $\mu = 6,7,8,9$, have a
half-integer mode expansion.  Their superconformal partners
$\psi^\mu$ then have an integer mode expansion and the ground state is
a representation of the corresponding Clifford algebra.  The vertex
operator is thus a spin field: the periodic $\psi^\mu$ contribute a
factor $V = e^{i ( H_3 + H_4)/2}$, where $H_{3,4}$ are
from the bosonization of the four periodic $\psi^{6,7,8,9}$~\cite{fms}.
We need only consider this part of the vertex operator, as the rest
is the same as in the 99 string and so has $\Omega^2 = +1$.  Now,
the operator product of $V$ with itself (which is in the 55 or 99
sector) involves $e^{i ( H_3 + H_4)}$, which is the bosonization of
$(\psi^6 + i \psi^7)(\psi^8 + i \psi^9)$. This in turn is the vertex
operator for the state  $(\psi^6 + i \psi^7)_{-1/2}(\psi^8 + i
\psi^9)_{-1/2} |0\rangle$.  Finally we can deduce the $\Omega$
eigenvalue.  For $|0\rangle$ it is $+1$, because its vertex operator
is the identity, while each $\psi_{-1/2}$ contributes either $-i$ (for
a 99 string) or $+i$ (for a 55 string), for an overall $-1$.  That is,
the $\Omega$ eigenvalue of $V \cdot V$ is $-1$, so therefore is the
$\Omega^2$ eigenvalue of $V$.

Returning to eq.~(\ref{oh2}), in the 59 sector there is an extra $-1$
from the above argument.  Separate $\gamma_{\Omega}$ into a
block $\gamma_{\Omega,9}$ which acts on the 9-branes and a block
$\gamma_{\Omega,5}$ which acts on the 5-branes.  We have
$\gamma_{\Omega,9}^T = +
\gamma_{\Omega,9}$ from tadpole cancellation.  To cancel the sign in the 59
sector we then need $\gamma_{\Omega,5}^T = - \gamma_{\Omega,5}$, giving
symplectic groups on the 5-brane as found in ref.~\cite{witinst}.
This argument seems roundabout, but it
is faithful to the logic that the actions of $\Omega$ in the 55 and 99
sectors are related because they are both contained in the 59 $\cdot$ 95
product.  Further, there does not appear to be any arbitrariness in
the result.

Let us briefly review the consequences of this
projection~\cite{witinst}.  Consider a pair of coincident 5-branes,
since the symplectic projection requires an even number.  The
world-brane vectors $\partial_t X^\mu$ ($\mu$ parallel to the 5-brane) have
Chan-Paton wavefunctions $\sigma^a_{ij}$, gauge group $USp(2) =
SU(2)$.  The world-brane scalars $\partial_n X^\mu$ ($\mu$ perpendicular to
the 5-brane) have Chan-Paton wavefunction $\delta_{ij}$.  Since these
are the collective coordinates for 5-branes~\cite{dlp}, the
wavefunction $\delta_{ij}$ means that the two
5-branes move together as a unit.  The need for this can also be seen
in another way~\cite{witpriv}.  In the Type~I theory the force
between 5-branes, and between 1-branes, is half of that calculated in
ref.~\cite{pold} because of the orientation projection.  The product
of the charges of a single 1-brane and single 5-brane would then be
only half a Dirac-Teitelboim-Nepomechie unit; but since the 5-branes
are always paired the quantization condition is respected.

The IIB theory also contains $p = 1$, 3, and 7-branes.  The above
argument gives $\Omega^2 = (\pm i)^{(9-p)/2}$.  This requires an
$SO$ projection on the 1-brane, consistent with Type I-heterotic
duality.  On the 3- and 7-branes it leads to an inconsistency.  This
is a satisfying result, as there is no conserved charge in the Type I
theory to give rise to such $p$-branes.

We do not know that we have found the complete set of consistency
conditions of this type, but no others are evident to us.

\subs{Tadpoles}

Modular invariance on the torus is one of the central consistency
requirements for closed oriented strings.  For open and unoriented
one-loop graphs there is no corresponding modular group, but
cancellation of divergences plays an analogous role in constraining the
theory~\cite{gsdiv}.  In refs.~\cite{cal1,cai} these divergences were
obtained in the ten-dimensional
Type~I theory from one-loop vacuum amplitudes.  In ref.~\cite{cai}
they were reinterpreted in terms of an inconsistency in the field
equation for a Ramond-Ramond (RR) 10-form potential.  It is useful to
recall the latter interpretation, now generalized to all RR forms. 
D-branes and orientifold fixed-planes are electric and magnetic
sources for the RR fields~\cite{pold}.  The $n$-form field strength
$H_n$ thus satisfies
\begin{equation} 
dH_n = *J_{9-n}, \qquad d*H_n = *J_{n-1} \label{fe}
\end{equation}
where $J_{9-n}$ and $J_{n-1}$ are sources of the indicated rank.
The field equations are consistent only if
\begin{equation}
\int_{C_{k}} *J_{10-k} = 0
\end{equation}
for all closed curves $C_k$.  In flat $d=10$ the only nontrivial
closed curves are the points $C_0$, and the corresponding constraint
on $J_{10}$ requires the gauge group $SO(32)$.  In a compact theory
there will be more constraints.

More generally, the right-hand side of the field equations~(\ref{fe}) will
include additional terms from Chern-Simons couplings of the RR
fields to curvature and gauge field strengths. In the present
work we consider only orientifolds of flat backgrounds, but the more
general case will also be interesting. 

The tadpole constraints were applied
to orientifolds in refs.~\cite{hor1,isho,bd,sag2}.  Many of the
results in the present section can be found in ref.~\cite{hor1},
except that our treatment of the Chan-Paton factors will be more
general.

\subsubsection{General Framework}

The divergences can be determined from the vacuum amplitudes on the
Klein bottle (KB), M\"obius strip (MS), and cylinder (C).  In fig.~1 these
surfaces are all depicted as cylinders of length~$2\pi l$ and circumference
$2\pi$, with the ends being either boundaries or crosscaps.
\begin{figure}
\begin{center}
\leavevmode
\epsfbox{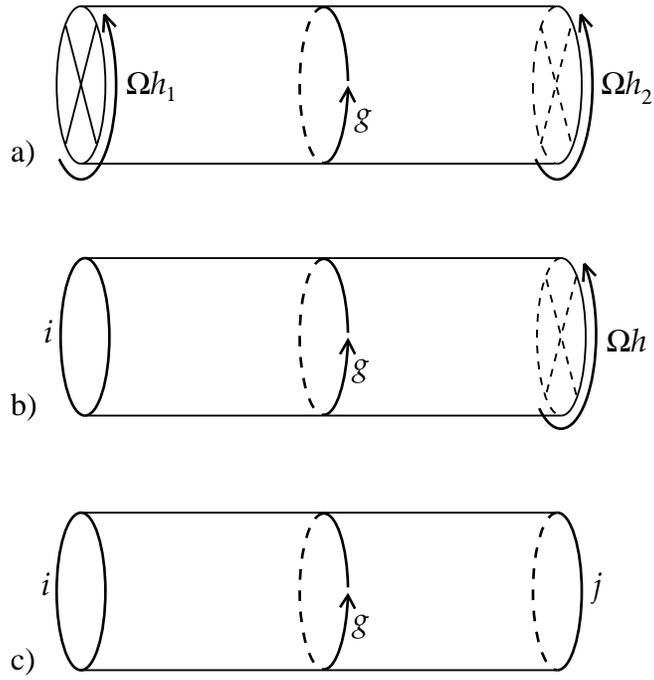}
\end{center}
\caption[]{Riemann surfaces described by eq.~(\ref{cond1}).
a) Klein bottle.  b) M\"obius strip. c) Cylinder.}
\end{figure}
Taking
coordinates $0 \leq \sigma^1 \leq 2\pi l$, $0 \leq \sigma^2 \leq 2\pi$, the
periodicity and boundary conditions on generic world-sheet
fields $\phi$ (and their derivatives) are as follows:
\begin{eqnarray}
\mbox{KB:}&& \phi(0,\pi+\sigma^2) = \Omega \tilde h_1 \phi(0,\sigma^2),
\quad \phi(2\pi l,\pi+\sigma^2) = \Omega\tilde h_2 \phi(2\pi
l,\sigma^2)\nonumber\\
&& \phi(\sigma^1,2\pi + \sigma^2) = \tilde
g\phi(\sigma^1,\sigma^2)\nonumber\\[4pt]
\mbox{MS:}&& \phi(0,\sigma^2) \in \tilde M_i,
\quad \phi(2\pi l,\pi+\sigma^2) = \Omega  \tilde h \phi(2\pi
l,\sigma^2)\nonumber\\ && \phi(\sigma^1,2\pi + \sigma^2) = \tilde
g\phi(\sigma^1,\sigma^2)\nonumber\\[4pt]
\mbox{C:}&& \phi(0,\sigma^2) \in \tilde M_i,
\quad \phi(2\pi,\sigma^2) \in \tilde M_j,\nonumber\\
&& \phi(\sigma^1,2\pi + \sigma^2) = \tilde g\phi(\sigma^1,\sigma^2). 
\label{cond1}
\end{eqnarray}
It is convenient to include in the periodicity or boundary conditions $g$, $h$,
and $M_i$, besides the spacetime part discussed earlier, a $\pm 1$ on the
world-sheet fermions from the GSO projection; the tilde is a reminder of this
additional information.
The respective definitions~(\ref{cond1}) are consistent only if
\begin{eqnarray}
\mbox{KB:}&& (\Omega \tilde h_1)^2 = (\Omega \tilde h_2)^2 = \tilde g
\nonumber\\ 
\mbox{MS:}&& (\Omega \tilde h)^2 = \tilde g, \quad \tilde g \tilde
M_i = \tilde M_i
\nonumber\\
\mbox{C:}&& \tilde g \tilde M_i = \tilde M_i, \quad \tilde g \tilde M_j =
\tilde M_j\ ;
\end{eqnarray}
else the corresponding path integral vanishes.

These graphs will have divergences from the tadpoles shown in fig.~2.
\begin{figure}
\begin{center}
\leavevmode
\epsfbox{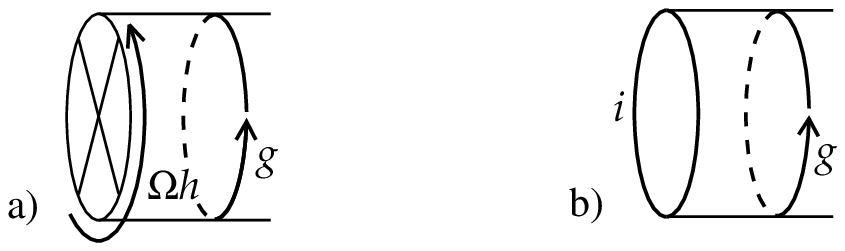}
\end{center}
\caption[]{Tadpoles in the $g$-twisted sector.
a) Crosscap: fields at opposite points differ by an $\Omega
h$ transformation, where $g = (\Omega h)^2$.  b) Boundary in state $i$. 
The manifold
$M_i$ must be fixed under $g$.} 
\end{figure} 
If there are $m$ non-compact dimensions, the
dangerous tadpoles will be from those massless RR states which are
$m$-forms in the non-compact directions. In general there are several
such tadpoles, coming from twisted and untwisted sectors.

To write the Klein bottle and M\"obius strips in terms of traces,
take the alternate coordinate region
$0 \leq \sigma^1 \leq 4\pi l$, $0 \leq \sigma^2 \leq \pi$ with
periodicities\footnote
{This is done by taking the upper strip $\pi \leq \sigma^2 \leq 2\pi$,
inverting it from right to left and multiplying the fields by $(\Omega
\tilde h_2)^{-1}$, and gluing it to the right side of the lower strip: with
this construction the fields are smooth at $\sigma^1 = 2\pi l$.}
\begin{eqnarray}
\mbox{KB:}&& \phi(\sigma^1,\pi+\sigma^2) = \Omega \tilde h_2 \phi(4\pi l
-\sigma^1,\sigma^2), \quad \phi(4\pi l,\sigma^2) = \tilde g'\phi(0,\sigma^2)
\nonumber\\
\mbox{MS:}&& \phi(\sigma^1,\pi+\sigma^2) = \Omega \tilde h \phi(4\pi l
-\sigma^1,\sigma^2), \quad \phi(0,\sigma^2) \in \tilde M_i,\nonumber\\
&& \qquad\qquad \phi(4\pi l,\sigma^2) \in \tilde M_i
\end{eqnarray}
where $\tilde g' = \Omega \tilde h_2 (\Omega \tilde h_1)^{-1}$.
Rescaling the coordinates to standard length ($\pi$ for open strings and
$2\pi$ for closed), the respective amplitudes are
\begin{eqnarray}
\mbox{KB:}&& {\rm Tr}_{{\rm c},g'} \left(\Omega \tilde h_2 (-1)^{F + \tilde
F} e^{\pi (L_0 + \tilde L_0)/2l} \right)
\nonumber\\
\mbox{MS:}&& {\rm Tr}_{{\rm o},ii} \left(\Omega \tilde h (-1)^{F}
e^{\pi L_0 /4l}\right)
\nonumber\\
\mbox{C:}&& {\rm Tr}_{{\rm o},ij} \left(\tilde g (-1)^{F}
e^{\pi L_0 /l}\right) \label{traces}
\end{eqnarray}
The closed string trace is labeled by the spacelike twist $g'$ and the
open string traces are labeled by the Chan-Paton states.

The full one-loop amplitude is
\begin{equation}
\int_0^\infty \frac{dt}{2t} 
\left\{ {\rm Tr}_{{\rm c}} \left({\bf P} (-1)^{\bf F} e^{-2\pi t(L_0 +
\tilde L_0)}\right) + {\rm Tr}_{{\rm o}} \left({\bf P} (-1)^{\bf F} e^{-2\pi
t L_0}\right)
\right\},
\end{equation}
where {\bf P} includes the GSO and $G$-projections, and {\bf F} is the
spacetime fermion number.  The traces are over the transverse oscillator
states and include a spacetime momentum sum.  The sums in the projection
operators and over twisted sectors and Chan-Paton states are equivalent
to summing the surfaces in fig.~1 over all tadpole types.  Evaluating
the traces, the $t \to 0$ limit produces the divergences of interest.
Note that the loop modulus $t$ is related to the cylinder length $l$
differently for each surface,
\begin{equation}
\mbox{KB:}\ \ t = \frac{1}{4l},\qquad
\mbox{MS:}\ \ t = \frac{1}{8l},\qquad
\mbox{C:}\ \ t = \frac{1}{2l}. \label{modrel}
\end{equation}

\subsubsection{Type I Theory on a $K3$ Orbifold}

We will evaluate the tadpoles and solve the consistency conditions for
one particular example. 
This is the Type I theory on a K3 $Z_2$ orbifold.  The Type I theory
includes a projection on $\Omega$.  The orbifold is formed from the
theory on a torus by projecting with $R_{6789}$, reflection of
$X^{6,7,8,9}$; we will abbreviate this as $R$.  Closure gives
also the element $\Omega R$.  To define $R$ we have to make a
specific choice of its action on the fermions; we choose
$R=e^{i\pi(J_{67}+J_{89})}$.

This example is of interest for a number of reasons.  It is
related by $T$-duality to many similar theories.  A $T$-duality
transformation on $X^\mu$ for given $\mu$ (abbreviated $T_\mu$) is a
spacetime reflection, but only on the world-sheet right-movers.  It
transforms $\Omega$ to  $\Omega R_\mu$~\cite{dlp,hor2}.  Thus,
$T_6$-duality takes the above orientifold group to $\{ 1,
R_{6789}, \Omega R_6, \Omega R_{789} \}$, $T_{4567}$-duality
takes it to $\{ 1, R_{6789}, \Omega R_{4567}, \Omega
R_{4589} \}$, and so on.  This is the simplest orientifold that is
not just the $T$-dual of a toroidal theory.

We can anticipate some of the tadpole calculation.  The
$\Omega$ projection will require 9-branes as in the non-compact
case~\cite{cai}.  Similarly the $\Omega R$ projection, being
$T_{6789}$-dual to $\Omega$, will require 5-branes with fixed
$X^{6,7,8,9}$.  There is also the possibility of twisted sector
tadpoles, and these will indeed appear.  In all there are three
tadpole types, the 10-form, 6-form, and twisted-sector 6-form
(actually 16 of these last, one for each fixed point) and each
receives two contributions.  The 10-form receives contributions from
the crosscap with $h=1$ and the 9-brane boundary with $g=1$, the
6-form from the crosscap with $h=R$ and the 5-brane boundary with
$g=1$, and the twisted-sector 6-forms from 5-brane and 9-brane
boundaries with $g = R$.

The IIB theory has $d=10, N=2 \to d=6, N=4$ spacetime supersymmetry.  The
$\Omega$ projection leaves only the sum of the left- and right-moving
supersymmetries, $Q_\alpha + \tilde Q_\alpha$.  Similarly the $\Omega R$
projection leaves the linear combinations $Q_\alpha + R \tilde Q_\alpha$.
The supersymmetries unbroken by both projections correspond to the $+1$
eigenvalues of $R$; this is half the eigenvalues of $R$ or a quarter of the
original supersymmetries, $d=6, N=1$.

Let us work out the massless spectrum of this theory.  We focus on the
bosons, since the fermions will have the same partition function by
supersymmetry. The massless spectrum for the right- or left-moving half of
the closed string is
\begin{equation}
\begin{array}{rlcc}
\mbox{Sector$\qquad$}&&R&SO(4)\mbox{ rep.}\\[4pt]
\mbox{Untwisted NS:\ \ }&\psi_{-1/2}^\mu|0\rangle &+&({\bf 2},{\bf 2})\\
&\psi_{-1/2}^m|0\rangle &-&4({\bf 1},{\bf 1})\\
\mbox{R:\ \ }&|s_1s_2s_3s_4\rangle &&\\
&\quad s_1=+s_2,\ s_3=-s_4\ &+&2({\bf 2},{\bf 1})\\
&\quad s_1=-s_2,\ s_3=+s_4\ &-&2({\bf 1},{\bf 2})\\
\mbox{Twisted NS:\ \ }&|s_3s_4\rangle,\ s_3 = -s_4&+&2({\bf 1},{\bf
1})\\
\mbox{R:\ \ }&|s_1s_2\rangle,\ s_1 = -s_2&+&({\bf 1},{\bf 2})\\
\end{array}
\end{equation}
Here, $\mu \in 2,3,4,5$, $m \in 6,7,8,9$, and $SO(4)=SU(2)\times SU(2)$ is
the massless little group in 6 dimensions.  We have imposed the GSO projection: all
states listed have $(-1)^F= (-1)^{\tilde F} = 1$.  This is most easily
determined by requiring that the vertex operators be local with respect
to the supercharge $e^{-\phi/2}e^{i(H_0+H_1+H_2+H_3-H_4)/2}$ (the minus sign
in the exponent is necessary because this must have $R=+$);
the ghost
times longitudinal part contributes a net $Z_2$ branch cut in the NS
sector and none in the R sector.  The bosonic
spectrum is given by the product a left-moving state with a right-moving
state from the same sector and with the same $R$.  In the NSNS sectors this
is symmetrized by the
$\Omega$ projection, and in the RR sectors it is antisymmetrized because
each side is a fermion.  Thus, including the degeneracy from the 16 fixed
points, the massless closed string spectrum is
\begin{equation}
\begin{array}{rc}
\mbox{Sector$\qquad$}&SO(4)\mbox{ rep.}\\[4pt]
\mbox{Untwisted NSNS: }& ({\bf 3},{\bf 3})+11 ({\bf 1},{\bf 1})\\
\mbox{RR: }& ({\bf 3},{\bf 1})+({\bf 1},{\bf 3})+6 ({\bf 1},{\bf
1})\\
\mbox{Twisted NSNS: }& 48({\bf 1},{\bf 1})\\
\mbox{RR: }&16({\bf 1},{\bf 1})\\
\end{array}
\end{equation}
This is the $d=6$, $N=1$ supergravity multiplet, plus one tensor
multiplet, plus 20 $({\bf 1},{\bf 1})$ hypermultiplets.

For the open strings consider first the 99 states, with massless bosonic
(NS) spectrum
\begin{equation}
\begin{array}{lllc}
&R=+&\Omega = +&SO(4)\mbox{ rep.}\\[4pt]
\psi_{-1/2}^\mu|0,ij\rangle \lambda_{ji}\ &
\lambda = \gamma_{R,9}\lambda \gamma_{R,9}^{-1} &\lambda =
-\gamma_{\Omega,9}\lambda^T
\gamma_{\Omega,9}^{-1}&({\bf 2},{\bf 2})\\ 
\psi_{-1/2}^m |0,ij\rangle \lambda_{ji}\ &
\lambda = -\gamma_{R,9} \lambda \gamma_{R,9}^{-1} &\lambda =
-\gamma_{\Omega,9}\lambda^T
\gamma_{\Omega,9}^{-1}&4({\bf 1},{\bf 1}).\\ 
\end{array}\label{os1}
\end{equation}
We have indicated the conditions imposed by the $R$ and $\Omega$
projections on the Chan-Paton wavefunctions $\lambda$.  The
subscript $9$ indicates the block of $\gamma_R$ or $\gamma_{\Omega}$ which
acts on the 9-branes. For the 55 open strings, let us first consider $n_I$
5-branes at the $I$'th fixed point of $R$.  The massless spectrum is
\begin{equation}
\begin{array}{lllc}
&R=+&\Omega = +&SO(4)\mbox{ rep.}\\[4pt]
\psi_{-1/2}^\mu|0,ij\rangle \lambda_{ji}\ &
\lambda = \gamma_{R,I} \lambda \gamma_{R,I}^{-1} &\lambda =
-\gamma_{\Omega,I}\lambda^T
\gamma_{\Omega,I}^{-1}&({\bf 2},{\bf 2})\\ 
\psi_{-1/2}^m |0,ij\rangle \lambda_{ji}\ &
\lambda = -\gamma_{R,I} \lambda \gamma_{R,I}^{-1} &\lambda =
\gamma_{\Omega,I}\lambda^T
\gamma_{\Omega,I}^{-1}&4({\bf 1},{\bf 1}),\\
\end{array} \label{os2}
\end{equation}
where now $\gamma_{R,I}$ and $\gamma_{\Omega,I}$ are the blocks acting on
this set of 5-branes.  The extra sign in the $\Omega$ projection follows from
the form of the vertex operator, as discussed earlier.  Now consider $n'_J$
5-branes at a non-fixed point $X$, which requires also $n'_J$ at $-X$.  The
massless bosonic strings with both ends at $X$ are
\begin{equation}
\begin{array}{llc}
&\Omega = +&SO(4)\mbox{ rep.}\\[4pt]
\psi_{-1/2}^\mu|0,ij\rangle \lambda_{ji}\quad&
\lambda = -\gamma'_{\Omega,J}\lambda^T
\gamma'^{-1}_{\Omega,J}\ &({\bf 2},{\bf 2})\\ 
\psi_{-1/2}^m |0,ij\rangle \lambda_{ji}\quad&
\lambda = \gamma'_{\Omega,J}\lambda^T
\gamma'^{-1}_{\Omega,J}\ &4({\bf 1},{\bf 1}).\\ 
\end{array} \label{os3}
\end{equation}
The $R$ projection relates these wavefunctions to those of the strings
with ends at $-X$, but does not otherwise constrain them.  For the 59
strings, we have in the two cases above
\begin{equation}
\begin{array}{llc}
&R = +&SO(4)\mbox{ rep.}\\[4pt]
|s_3 s_4,ij\rangle \lambda_{ji},\ s_3 = -s_4\quad&
\lambda = \gamma_{R,I}\lambda \gamma_{R,9}^{-1}\ &2({\bf 1},{\bf 1})\\ 
\end{array}\label{os4}
\end{equation}
and
\begin{equation}
\begin{array}{lc}
&SO(4)\mbox{ rep.}\\[4pt]
|s_3 s_4,ij\rangle \lambda_{ji},\ s_3 = -s_4\qquad& 2({\bf 1},{\bf 1}).\\
\end{array} \label{os5}
\end{equation}
The $\Omega$ projection does not constrain these but determines the 95
state in terms of the 59 states.

\subsubsection{Tadpole Calculation}

We may now evaluate the sums~(\ref{traces}) over the closed and open
string spectra.  The amplitudes are $\int_0^\infty dt/2t$ times
\begin{eqnarray} 
\mbox{KB:} && {\rm Tr}_{\rm NSNS + RR}^{\rm U+T}
\left\{ \frac{\Omega}{2}\,\frac{1+R}{2}\,\frac{1+(-1)^{F}}{2}
e^{-2\pi t(L_0 + \tilde L_0)}\right\} \nonumber\\
\mbox{MS:} && {\rm Tr}_{\rm NS-R}^{99+55}
\left\{ \frac{\Omega}{2}\,\frac{1+R}{2}\,\frac{1+(-1)^{F}}{2}
e^{-2\pi t L_0}\right\} \nonumber\\
\mbox{C:} && {\rm Tr}_{\rm NS-R}^{99+95+59+55}
\left\{ \frac{1}{2}\,\frac{1+R}{2}\,\frac{1+(-1)^{F}}{2} 
e^{-2\pi t L_0}\right\} . \label{amps}
\end{eqnarray} 
Here U (T) refers to the untwisted (twisted) closed string sector.  On the
Klein bottle we omit the $\frac{1}{2}(1+(-1)^{\tilde{F}})$ projector
because the left- and  right-moving states are identical in the trace.
The open-string traces include a sum over Chan-Paton states.

The signs of the operators appearing in the traces, in
the various sectors, were given implicitly in the previous section.  For
completeness we give the action of $\Omega$ on the various mode operators;
the action of $R$ is obvious.  In the closed string,
\begin{equation}
\Omega \alpha_r \Omega^{-1} = \tilde \alpha_r, \qquad
\Omega \psi_r \Omega^{-1} = \tilde \psi_r, \qquad
\Omega \tilde\psi_r \Omega^{-1} = - \psi_r
\end{equation}
for integer and half-integer $r$.  The minus sign is included in the last
equation to give the convenient result $\Omega \psi_M \tilde\psi_M
\Omega^{-1} = \psi_M \tilde\psi_M$ for any product $\psi_M$ of mode
operators.  Alternately this sign can be omitted: this just corresponds to
$\Omega \to (-1)^F \Omega$, which has the same action on physical states.
In open string, the mode expansions are
\begin{equation}
X(\sigma,0) = x + i
\sqrt{\frac{\alpha'}{2} } \sum_{ { m={-\infty} }
\atop {m \neq 0} }^\infty
\frac{\alpha_m}{m} ( e^{im\sigma} \pm e^{-im\sigma} ) 
\end{equation}
with the upper sign for NN boundaries conditions and lower for DD
(N=Neu\-mann, D=Dirichlet).  World-sheet parity, $X(\sigma,0) \to
X(\pi-\sigma,0)$, takes
\begin{equation}
\alpha_m \to \pm e^{i\pi m} \alpha_{m}.
\end{equation}
There is no corresponding result for the ND sector, since $\Omega$ takes
this into a different, DN, sector.
For fermions, the mode expansions are
\begin{equation}
\psi(\sigma,0) = \sum_r e^{ir\sigma} \psi_r, \qquad  
\tilde\psi(\sigma,0) = \sum_r e^{-ir\sigma} \psi_r.
\end{equation}
Parity, $\psi(\sigma,0) \to \pm\tilde\psi(\pi-\sigma,0)$, takes
\begin{equation}
\psi_r \to \pm e^{i\pi r} \psi_{r}
\end{equation}
for integer and half-integer $r$.
As in the closed string there is some physically irrelevant sign freedom.
In evaluating the traces, note that $\Omega$ and $R$ act on the compact
momenta $p^m$ and windings $\delta X^m = X^m(2\pi) - X^m(0)$ of the 
closed
string as \begin{eqnarray}
&& \Omega p^m \Omega^{-1} = p^m, \qquad Rp^mR^{-1} = -p^m \nonumber\\
&& \Omega \delta X^m \Omega^{-1} = -\delta X^m, \qquad R\delta X^m
R^{-1} = -\delta X^m 
\end{eqnarray}
and that only diagonal elements contribute in the traces.  Similarly in the
open string 99 sector there is an internal momentum, while in the 55
sector with fixed endpoints there is a winding $\delta X^m = X^m(\pi) -
X^m(0)$.

It is useful to define\footnote{The $\sqrt 2$ in $f_2$ corrects a
typographical error in ref.~\cite{cai}.}
\begin{eqnarray}
f_{1}(q) = q^{1/12} \prod_{n=1}^\infty \left(1-q^{2n}\right),\qquad
f_{2}(q) = q^{1/12} \sqrt{2}\,\prod_{n=1}^\infty \left(1+q^{2n}\right)
\nonumber\\ 
f_{3}(q) = q^{-1/24} \prod_{n=1}^\infty \left(1+q^{2n-1}\right),\qquad
f_{4}(q) = q^{-1/24}\prod_{n=1}^\infty \left(1-q^{2n-1}\right) .
\end{eqnarray}
These functions satisfy the `abstruse identity'
\begin{eqnarray}f_{3}^{8}(q) = f_{2}^{8}(q)+f_{4}^{8}(q)\end{eqnarray}
and have the modular transformations
\begin{equation}
f_{1}(e^{-{\pi}/{s}}) = \sqrt{s}\,f_{1}(e^{-\pi s}),\quad
f_{3}(e^{-{\pi}/{s}}) = f_{3}(e^{-\pi s}),\quad
f_{2}(e^{-{\pi}/{s}}) = f_{4}(e^{-\pi s}) . \label{modt}
\end{equation}

The amplitudes~(\ref{amps}), including the integrals over non-compact
momenta, are then found to be
$
(1-1)\frac{v_6}{128} \int_0^\infty \frac{dt}{t^4}
$
times
\begin{eqnarray}
\mbox{KB:}&&   
8\frac{f_{4}^{8}(e^{-2\pi t})}{f_{1}^{8}(e^{-2\pi t})}
\left\{ \left( \sum_{n=-\infty}^\infty e^{-\pi t n^{2}/\rho} \right)^{4}
+ \left( \sum_{w=-\infty}^\infty e^{-\pi t \rho w^{2}} \right)^{4}
\right\} \nonumber\\[4pt]
\mbox{MS:} && - \frac{f_{2}^{8}(e^{-2\pi t})f_{4}^{8}(e^{-2\pi t})}
{f_{1}^{8}(e^{-2\pi t})f_{3}^{8}(e^{-2\pi t})}
\left\{ {\rm Tr}(\gamma_{\Omega,9}^{-1}\gamma_{\Omega,9}^T)
\left(\sum_{n=-\infty}^\infty e^{- 2\pi t n^{2}/\rho} \right)^{4} 
\right.\nonumber\\
&&\left. \qquad+ {\rm Tr}(\gamma_{\Omega R,5}^{-1}\gamma_{\Omega R,5}^T) 
\left(\sum_{w=-\infty}^\infty 
e^{-2\pi t\rho w^{2}} \right)^{4} \right\}
\nonumber\\[4pt]
\mbox{C:}&& \frac{f_{4}^{8}(e^{-\pi t})}{f_{1}^{8}(e^{-\pi t})} 
\left\{ ({\rm Tr}(\gamma_{1,9}))^2
\left(\sum_{n=-\infty}^\infty e^{- 2\pi t n^{2}/\rho} \right)^{4}
 \right.\nonumber\\
&& \left.\qquad + \sum_{i,j\in 5} (\gamma_{1,5})_{ii} (\gamma_{1,5})_{jj}
\prod_{m=6}^9 \sum_{w=-\infty}^\infty 
e^{-t (2\pi wr +X^m_i - X^m_j)^{2}/2\pi\alpha'} \right\}\nonumber\\
&-& 2  \frac{f_{2}^{4}(e^{-\pi t})f_{4}^{4}(e^{-\pi t})}
{f_{1}^{4}(e^{-\pi t})f_{3}^{4}(e^{-\pi t})} {\rm Tr}(\gamma_{R,5})
{\rm Tr}(\gamma_{R,9}) \nonumber\\
&+& 4 \frac{f_{3}^{4}(e^{-\pi t})f_{4}^{4}(e^{-\pi t})}
{f_{1}^{4}(e^{-\pi t})f_{2}^{4}(e^{-\pi t})} 
\left\{ ({\rm Tr}(\gamma_{R,9}))^2 + \sum_{I=1}^{16} 
({\rm Tr}(\gamma_{R,I}))^2 \right\} .
\end{eqnarray} 
We have defined $v_6 = V_6 /(4\pi^2 \alpha')^3$ where $V_6$ is the
(regulated) volume of the non-compact dimensions.  Also, $2\pi r$ is the
periodicity of $X^m$ (assumed for convenience to be
independent of $m$), $\rho = r^2/\alpha'$, and we will later use $v_4 =
\rho^2 = V_4/(4\pi^2 \alpha')^2$ with $V_4$ the volume of the torus
before the orientifold.  The second term in the first brace of the
cylinder amplitude includes a sum over 5-brane pairs $M_i M_j$ and over all
ways for an open string to wind from one to  the other.  In the second term
in the second brace of the cylinder, the only diagonal elements of $R$ are
those where the open string begins and ends on the same 5-brane without
winding---hence the sum over fixed points $I$.

Using the modular transformation~(\ref{modt}) and the Poisson
resummation formula
\begin{equation}
\sum_{n=-\infty}^{\infty} e^{-\pi (n-b)^2/a} =
\sqrt{a} \sum_{s=-\infty}^{\infty} e^{-\pi a s^2 + 2\pi i s b},
\end{equation}
the amplitude becomes
$
(1-1) \int_0^\infty \frac{dt}{t^2}
$
times
\begin{eqnarray}
\mbox{KB:}&&   
\frac{f_{2}^{8}(e^{-\pi/2 t})}{f_{1}^{8}(e^{-\pi/2 t})}
\left\{ v_6 v_4 \left( \sum_{s=-\infty}^\infty e^{-\pi \rho s^{2}/t}
\right)^{4} + \frac{v_6}{v_4} \left( \sum_{s=-\infty}^\infty e^{-\pi s^2/ t
\rho} \right)^{4} \right\} \nonumber\\[4pt]
\mbox{MS:} && -\frac{f_{2}^{8}(e^{-\pi/ 2t})f_{4}^{8}(e^{-\pi /2t})}
{f_{1}^{8}(e^{-\pi/ 2t})f_{3}^{8}(e^{-\pi/ 2t})}
\left\{ \frac{v_6v_4}{32} {\rm Tr}(\gamma_{\Omega,9}^{-1}\gamma_{\Omega,9}^T)
\left(\sum_{s=-\infty}^\infty e^{- \pi \rho s^{2}/2t}
\right)^{4} 
\right.\nonumber\\
&&\left. \qquad+ \frac{v_6}{32 v_4} {\rm
Tr}(\gamma_{\Omega R,5}^{-1}\gamma_{\Omega R,5}^T) 
\left(\sum_{s=-\infty}^\infty 
e^{-\pi s^{2}/ 2t\rho} \right)^{4} \right\}
\nonumber\\[4pt]
\mbox{C:}&& \frac{f_{2}^{8}(e^{-\pi /t})}{f_{1}^{8}(e^{-\pi /t})} 
\left\{ \frac{v_6v_4}{512} ({\rm Tr}(\gamma_{1,9}))^2
\left(\sum_{s=-\infty}^\infty e^{- \pi\rho s^{2}/2t}
\right)^{4} \right.\nonumber\\
&& \left.\qquad + \frac{v_6}{512 v_4} 
\sum_{i,j\in 5} (\gamma_{1,5})_{ii} (\gamma_{1,5})_{jj}
\prod_{m=6}^9 \sum_{s=-\infty}^\infty 
e^{-\pi s^2/2t\rho + is (X^m_i - X^m_j)/r}
\right\}\nonumber\\ 
&-& \frac{f_{2}^{4}(e^{-\pi/ t})f_{4}^{4}(e^{-\pi
/t})} {f_{1}^{4}(e^{-\pi /t})f_{3}^{4}(e^{-\pi /t})} 
\frac{v_6}{64} {\rm Tr}(\gamma_{R,5})
{\rm Tr}(\gamma_{R,9}) \nonumber\\
&+& \frac{f_{3}^{4}(e^{-\pi /t})f_{2}^{4}(e^{-\pi /t})}
{f_{1}^{4}(e^{-\pi /t})f_{4}^{4}(e^{-\pi /t})} 
\frac{v_6}{32} \left\{ ({\rm Tr}(\gamma_{R,9}))^2 + \sum_{I=1}^{16} 
({\rm Tr}(\gamma_{R,I}))^2 \right\} .
\end{eqnarray} 
The asymptotics are
\begin{eqnarray}
\mbox{KB:}&&   
16 v_6 v_4  + 16 \frac{v_6}{v_4}  \nonumber\\[4pt]
\mbox{MS:} && -
\frac{v_6v_4}{2} {\rm Tr}(\gamma_{\Omega,9}^{-1}\gamma_{\Omega,9}^T)
- \frac{v_6}{2 v_4}  {\rm Tr}(\gamma_{\Omega R,5}^{-1}\gamma_{\Omega R,5}^T) 
\nonumber\\[4pt]
\mbox{C:}&& 
\frac{v_6v_4}{32} ({\rm Tr}(\gamma_{1,9}))^2+ \frac{v_6}{32 v_4} 
({\rm Tr}(\gamma_{1,5}))^2
\nonumber\\ 
&-& 
\frac{v_6}{16} {\rm Tr}(\gamma_{R,5})
{\rm Tr}(\gamma_{R,9}) \nonumber\\
&+& 
\frac{v_6}{8} \left\{ ({\rm Tr}(\gamma_{R,9}))^2 + \sum_{I=1}^{16} 
({\rm Tr}(\gamma_{R,I}))^2 \right\} .
\end{eqnarray} 

Finally, the total amplitude for large $l$ (noting the
relations~(\ref{modrel}) between $t$ and $l$) is $(1-1) \int_0^\infty dl$
times
\begin{eqnarray}
&&\frac{v_6 v_4}{16} \left\{ 32^2 - 64 {\rm
Tr}(\gamma_{\Omega,9}^{-1}\gamma_{\Omega,9}^T) + ({\rm Tr}(\gamma_{1,9}))^2
\right\}
\nonumber\\ &+& \frac{v_6}{16 v_4} \left\{ 32^2 - 64 {\rm
Tr}(\gamma_{\Omega R,5}^{-1}\gamma_{\Omega R,5}^T) + ({\rm
Tr}(\gamma_{1,5}))^2
\right\}
\nonumber\\ &+& \frac{v_6}{64} \sum_{I=1}^{16} \left({\rm
Tr}(\gamma_{R,9}) - 4{\rm Tr}(\gamma_{R,I}) \right)^2. \label{divs}
\end{eqnarray}
The $(1-1)$ represents the contributions of NSNS and RR exchange.  These
$l \to \infty$ divergences are
equal and opposite by supersymmetry, but must
vanish separately in a consistent theory~\cite{cai}.
The divergences have
the expected form.  The RR part of the first line,
proportional to the total spacetime volume, is from exchange of the
10-form; in the second line, proportional to the $T$-dual spacetime volume,
it is from exchange of the 6-form; in the third line, independent of the volume
of the internal spacetime, it is from exchange of twisted-sector 6-forms, one
for each fixed point.  Note that $\gamma_1 = 1$ and so ${\rm
Tr}(\gamma_{1,9}) = n_9$, ${\rm Tr}(\gamma_{1,5}) = n_5$, the numbers of
9-branes and 5-branes respectively.

\subs{Solutions}

Now let use solve the consistency conditions we have found, from the algebra
of Chan-Paton matrices and the cancellation of divergences.  From the
algebra we have eq.~(\ref{chitran}), implying $\gamma_{\Omega,9}^T = \pm
\gamma_{\Omega,9}$ and $\gamma_{\Omega R,5}^T = \pm' \gamma_{\Omega R,5}$. 
The 10-form and 6-form divergences are thus proportional to $(32 \mp n_9)^2$
and $(32 \mp' n_5)^2$, and so
\begin{eqnarray}
&& n_9 = 32, \qquad \gamma_{\Omega,9}^T = \gamma_{\Omega,9}, \qquad
\gamma_{\Omega,5}^T = -
\gamma_{\Omega,5} \nonumber\\
&& n_5 = 32, \qquad \gamma_{\Omega R,5}^T = \gamma_{\Omega R,5}, \qquad
\gamma_{\Omega R,9}^T = -
\gamma_{\Omega R,9} .
\end{eqnarray}
The last equality in each line follows from the discussion at the end of
section~2.  By a unitary change of basis, $\gamma_{\Omega h} \to U
\gamma_{\Omega h} U^T$, one can take 
\begin{equation}
\gamma_{\Omega,9} = I, \qquad \gamma_{\Omega R,5} = I.  \label{ident}
\end{equation}
The remaining constraints from the algebra are
\begin{eqnarray}
&& \gamma_{R,9} = \gamma_{\Omega,9} \gamma_{\Omega R,9} = \gamma_{\Omega R,9}
\nonumber\\ && \gamma_{R,5} = \gamma_{\Omega,5} \gamma_{\Omega R,5} =
\gamma_{\Omega,5}
\nonumber\\ && \gamma_{R,5}^2 = \gamma_{R,9}^2 = 1,
\end{eqnarray}
where all phases have been set to one by choice of the irrelevant overall
phases of $\gamma_{R,9}$, $\gamma_{R,5}$, $\gamma_{\Omega R,9}$, and
$\gamma_{\Omega,5}$. Together with the unitarity of these matrices, this
implies that they are Hermitean, as well as antisymmetric.  The
choice~(\ref{ident}) leaves the freedom to make real orthogonal
transformations.  With this, we can take
\begin{equation}
\gamma_{R,9} = \gamma_{R,5} = \gamma_{\Omega R,9} = \gamma_{\Omega,5}
= M = \left[ \begin{array}{cc} 0&iI\\-iI&0
\end{array} \right]\ , \label{ms}
\end{equation}
the blocks being $16 \times 16$.
Finally, the twisted sector tadpoles vanish.  Thus we have found a unique consistent
solution for the action of the symmetries on the Chan-Paton factors.

Returning to the massless spectra in section~3.2, we can now solve for the
Chan-Paton wavefunctions.  For the 99 open strings, eq.~(\ref{os1}) implies
the wavefunctions
\begin{eqnarray}
\mbox{vectors:}&&\lambda = \left[ \begin{array}{cc} A&S\\-S&A \end{array}
\right] \nonumber\\
\mbox{scalars:}&&\lambda = \left[ \begin{array}{cc} A_1&A_2\\A_2&-A_1
\end{array} \label{99wf}
\right]
\end{eqnarray}
where $S$ and $A$ refer to symmetric and antisymmetric blocks respectively.
The vectors form the adjoint of $U(16)$, with the Chan-Paton index
transforming as ${\bf 16} + {\bf \overline {16}}$.  The scalars transform as
the antisymmetric tensor ${\bf 120} + {\bf \overline {120}}$ of $U(16)$.  The
scalars are in sets of four, from
$m=6,7,8,9$, which is the content of a hypermultiplet.  Thus the 99 sector
contains a vector multiplet in the adjoint of $U(16)$ and hypermultiplets in
the ${\bf 120} + {\bf \overline {120}}$ (or equivalently, two hypermultiplets
in the ${\bf 120}$). 

For 55 open strings, consider first $n_I$ D-branes at
fixed point $I$; $n_I \equiv 2m_I$ must be even in order for the
matrices~(\ref{ms}) to have a sensible action.  For open strings with both
ends at $I$, eq.~(\ref{os2}) gives the same wavefunctions~(\ref{99wf}) as
for the 99 strings, a vector multiplet in the adjoint of
$U(m_I)$ and two hypermultiplets in the antisymmetric ${\bf\frac{1}{2}}
{\bf m}_I ({\bf m}_I {\bf -1})$ of this group. Now consider $n'_J$
D-branes at a non-fixed point
$X$, where again
$n'_J
\equiv 2m'_J$ must be even.  Eq.~(\ref{os3}) implies that the vector
multiplets are in the adjoint representation of $USp(n'_J)$ and the
hypermultiplets are in one antisymmetric ${\bf\frac{1}{2}} {\bf n}'_J
({\bf n}'_J {\bf -1})$ (which is reducible in $USp(n'_J)$, containing one
singlet state).

For 59 open strings with the 5-brane at a fixed point, eq.~(\ref{os4})
implies the wavefunctions
\begin{eqnarray}
\mbox{scalars:}&&\lambda = \left[ \begin{array}{cc} X_1&X_2\\-X_2&X_1
\end{array}
\right]
\end{eqnarray}
with $X_1$ and $X_2$ general $16 \times m_I$ matrices.  These states
transform as the $({\bf 16},{\bf m}_I) + ({\bf \overline {16}},{\bf \overline
m}_I)$ of $U(16) \times U(m_I)$, but because there are only two scalar
states~(\ref{os4}) in this sector, this is a single hypermultiplet in the
$({\bf 16},{\bf m}_I)$.  Similarly, for 59 strings with 5-brane
not at a fixed point, eq.~(\ref{os5}) gives a hypermultiplet in the
$({\bf 16},{\bf n}'_J)$.

The total gauge group is
\begin{equation}
U(16) \times \prod_{I=1}^{16} U(m_I) \times \prod_J USp(n'_J), \qquad
\sum_{I=1}^{16} m_I + \sum_J n'_J = 16,
\end{equation}
with hypermultiplets in the representations
\begin{eqnarray}
&&2\left({\bf 120},{\bf 1},{\bf 1}\right)\ 
+\ \sum_{I=1}^{16}\left\{ 2\left({\bf 1},\mbox{${\bf\frac{1}{2}}$} {\bf
m}_I ({\bf m}_I {\bf -1}),{\bf 1}\right)_I +  \left({\bf 16},{\bf
m}_I,{\bf 1}\right)_I \right\} \nonumber\\
&&+\ \sum_{J}\left\{ \left({\bf
1},{\bf 1},\mbox{${\bf\frac{1}{2}}$} {\bf n}'_J ({\bf n}'_J {\bf
-1})\bf{-1}\right)_J + ({\bf 1},{\bf 1},{\bf 1}) +
\left({\bf 16},{\bf 1},{\bf n}'_J\right)_J \right\}.
\end{eqnarray}
We have checked that the $R^4$ and $F^4$ anomalies cancel for this
spectrum.\footnote
{Though these are not among the anomaly-free theories recently found in
ref.~\cite{jsan}.}
Much of this space of
theories is connected.  A multiple of four 5-branes can move away from a
fixed point.  A single pair forms the basic dynamical 5-brane and must move
together as discussed in section~2, and in the orientifold an image pair
must move in the opposite direction.  If $4k$ 5-branes move away from
fixed point $I$,
$U(m_I)$ breaks to $U(m_I - 2k)
\times USp(2k)$.  The collective coordinate for this motion is one of the
antisymmetric tensors of $U(m_I)$, which can indeed break $U(m_I)$ in this
way. Because $m_I$ can change only mod 2, there are disconnected sectors of
moduli space according to whether each of the $m_I$ is odd or even.
The largest group is $U(16) \times U(16)$ with all 5-branes on a single
fixed point.  Incidentally, we implicitly began with a torus with no Wilson
lines, but these Wilson lines (transforming again as the antisymmetric tensor
of $U(16)$) can break the 99 $U(16)$ in the same pattern as the 55.

The spacetime anomalies of this model will be discussed further in a
future publication~\cite{alles}.  The spectrum above has $U(1)$
anomalies, which are cancelled by a generalization of the Green-Schwarz
mechanism.  This also generates masses of order $g_{\it s}^{1/2}$ for up
to 16 $U(1)$ multiplets, so the above spectrum is only correct in the
formal $g_{\it s} \to 0$ limit.

\subs{Discussion}

The surprise is that we have not found the theory that we most expected, the
$K3$ orbifold with spin connection embedded in the gauge connection.  This
has gauge group $SO(28) \times SU(2)$, possibly enhanced at 
special points.  This theory must exist because it exists
for Type I on a smooth $K3$, where the spectrum is the same as for the
heterotic string because the low energy supergravities are the same. 
The question is the nature of the orbifold limit. We believe that what
is happening is as follows.  For Type I on a smooth K3, there is only
one kind of endpoint, with Neumann boundary conditions.  As we
approach the orbifold limit, some wavefunctions become localized at
the fixed point while others remain extended.  In the limit, endpoints
in localized states become Dirichlet endpoints, while endpoints in the
extended states remain Neumann.  But there is no reason for a
transition from one type of state to the other to be forbidden,
particularly as we have neglected the coupling of the endpoint to the
rest of the string.  This would correspond to a term in the
world-sheet action which changes the boundary condition from 5-brane
to 9-brane, which is just a 59 open string vertex operator.  So we
conjecture that the orbifold limit is a theory with nontrivial 59
backgrounds.  This is no longer a free world-sheet theory.  In fact,
it is rather complicated, similar to an orbifold with a twisted-state
background.

Let us pursue this a bit further.  Embedding the spin connection in the
gauge connection means that $\gamma_{R,9} = {\rm diag}(+1^{28},-1^4)$.
Section~2 then implies that $\gamma_{R,5}$ is antisymmetric.  This makes it
impossible to cancel the 6-form tadpole in the theories we consider, but if
we ignore this for now we might guess that we still need 32 5-branes, two at
each fixed point.  This gives an $SU(2)$ at each fixed point, for total
gauge group $SO(28) \times SO(4) \times SU(2)^{16}$.  Now, we have noted in
the beginning of section~3 that open string field strengths are also sources
for RR fields.  So it may be that it is possible to cancel the tadpoles with
an appropriate 59 background.  Moreover, some 59 strings are in $({\bf
2},{\bf 2})$'s of $SO(4)$ and {\bf 2}'s of one of the fixed point $SU(2)$'s
making it possible to break down to a diagonal $SU(2)$ and obtain the
expected gauge group with hypermultiplet ${\bf 2}$'s (which exist on
the smooth $K3$ but cannot be obtained directly from the ${\bf 32}$ of
$SO(32)$).\footnote{It is interesting to ask whether the theories we
have found are in the same moduli space as the spin=gauge orbifold, 
with a different gauge background; we have not answered this.}

There has been a substantial literature on this and related models. 
Ref.~\cite{hm} considered the Type I theory on K3 with spin connection
embedded in gauge connection and did not find an anomaly-free theory based on
a free CFT.  This is consistent with the discussion above.  Further, they
were led to argue for Dirichlet open strings with $SU(2)$ Chan-Paton
factors at each fixed point, as above, and that these $SU(2)$'s should
be identified with an $SU(2)$ in $SO(32)$, again consistent with the
discussion of the 59 background.  However, we do not see much hope for
making this more precise, because of the complicated nature of the
world-sheet theory.

Refs.~\cite{hor1,isho} also considered Type I on K3, but without embedding
the spin connection in the gauge connection.  Both implicitly assumed diagonal
$\gamma$'s, and neither observed the need~\cite{witinst} for a symplectic
projection on the 5-branes.  Ref.~\cite{isho} found it impossible to cancel
all tadpoles, though we do not understand their calculation in detail. 
Ref.~\cite{hor2} found a model with group $SO(32) \times SO(2)^{16}$.
However, because of the orthogonal projection on the 5-branes, this suffers
from the problem observed in ref.~\cite{witinst}: there are
half-hypermultiplets in real (not pseudoreal) representations. Also,
it is not clear that the twisted tadpoles cancel for this model: they
are discussed but the relative normalization does not seem to agree
with our result~(\ref{divs}).  Refs.~\cite{sag2} find anomaly-free
orientifold models with gauge groups such as $USp(8)^4$ that also
arise in our construction, though the matter content is different (no
antisymmetric tensors of the gauge group, but additional antisymmetric
tensor supergravity multiplets).  These models are constructed from
more abstract CFT's, and the description is rather inexplicit, so we
have not been able to make a detailed comparison.  Also,
refs.~\cite{bd} construct similar models using free fermions. 
Curiously the gauge groups are smaller than those found elsewhere
(e.g. $USp(8)$ in $d=6$); again we have not been able to make detailed
comparisons.

As an aside, there is a strong temptation to regard the Dirichlet open
strings as the twisted sector that is otherwise absent for the
orientifold~\cite{hm,sag1,hor1}.  This is true in a number of formal
senses, but we find it somewhat dangerous to think this way, in that it
might lead one only to a subset of the consistent theories.  Note,
too, that it is not always true: a D-brane in a non-compact space need not
be accompanied by an orientifold---there is no inconsistency in the RR
field equation because the flux can escape to infinity.  Conversely an
orientifold of a non-compact space does not require the
introduction of D-branes.

In conclusion, we have developed some of the necessary technology of
orientifolds and D-manifolds, as a step toward trying to uncover the
structure underlying string duality.  It will be interesting to analyze
the duality symmetries of these theories~\cite{alles}.

\subsection*{Acknowledgments} 

We are grateful to Tom Banks, Shyamoli Chaudhuri, Michael Douglas, Rob
Leigh, Nati Seiberg, Andrew Strominger, and Edward Witten for helpful
discussions.  This work is supported by NSF grants PHY91-16964 and
PHY94-07194.

\end{document}